\documentclass[%
 reprint,
 groupedaddress,
 amsmath,
 amssymb,
 aps, prl,
]{revtex4-1}

\usepackage{graphicx}
\usepackage{bm}
\usepackage{natbib}

\usepackage{float}
\usepackage{etoolbox}

\usepackage{tabularx}
\usepackage{multirow}
\usepackage{makecell}

\newcommand{\beginsupplement}{%
        \setcounter{table}{0}
        \renewcommand{\thetable}{S\arabic{table}}%
        \setcounter{figure}{0}
        \renewcommand{\thefigure}{S\arabic{figure}}%
        \setcounter{equation}{0}
        \renewcommand{\theequation}{S\arabic{equation}}%
        
        \onecolumngrid
       	\patchcmd{\subsection}
          {\centering}
          {\raggedright}
          {}
          {}
      	\patchcmd{\subsubsection}
          {\centering}
          {\raggedright}
          {}
          {}
          
     }

\begin{document}

\preprint{APS/123-QED}

\title{Probing gravity by holding atoms for 20 seconds}

\author{Victoria Xu}
    \email{vxu@berkeley.edu}
\author{Matt Jaffe}
\author{Cristian D. Panda}
\author{Sofus L. Kristensen}
\altaffiliation[Now at: ]{Niels Bohr Institute, University of Copenhagen, DK-2100 Copenhagen, Denmark.}%
\author{Logan W. Clark}
\altaffiliation[Also at: ]{James Franck Institute and Department of Physics, University of Chicago, Chicago, IL, USA.}%
\author{Holger M{\"u}ller}
    \email{hm@berkeley.edu}
\affiliation{Department of Physics, University of California, Berkeley, California 94720, USA}

\date{\today}

\begin{abstract}
Atom interferometers are powerful tools for both measurements in fundamental physics and inertial sensing applications. Their performance, however, has been limited by the available interrogation time of freely falling atoms in a gravitational field. We realize an unprecedented interrogation time of 20 seconds by suspending the spatially-separated atomic wavepackets in a lattice formed by the mode of an optical cavity. Unlike traditional atom interferometers, this approach allows potentials to be measured by holding, rather than dropping, atoms. After seconds of hold time, gravitational potential energy differences from as little as microns of vertical separation generate megaradians of interferometer phase. This trapped geometry suppresses the phase sensitivity to vibrations by 3-4 orders of magnitude, overcoming the dominant noise source in atom-interferometric gravimeters. Finally, we study the wavefunction dynamics driven by gravitational potential gradients across neighboring lattice sites. 
\end{abstract}

\maketitle

Matter-wave interferometers with freely falling atoms have demonstrated the ability to precisely measure e.g., gravity \cite{Peters1999-rn} and fundamental constants \cite{Rosi2014-kt,Parker2018-mo}, to test general relativity \cite{Muller2010-ia,Schlippert2014-on,Zhou2015-dm}, and to search for new forces \cite{Jaffe2017-ow,Haslinger2018-iq}. A major obstacle to increasing their sensitivity, however, has been the limited time during which coherent, spatially-separated superpositions of atomic wave packets can be interrogated. Up to 2.3 seconds of interrogation time has been realized in a 10-meter atomic fountain \cite{Dickerson2013-yo}, and several seconds of interrogation time are the target of experiments in fountains measuring hundreds of meters \cite{Coleman2018-eh,Zhan2019-zu}, zero-gravity planes \cite{Barrett2016-or}, drop towers \cite{Muntinga2013-fi}, sounding rockets \cite{Becker2018-ch}, and the International Space Station \cite{Tino2013-fc,Aguilera2014-dk,Elliott2018-qc}. Geometries that trap the interferometer in an optical lattice \cite{Charriere2012-sa,Zhang2016-dl} have been explored, but attempts to date have suffered from dephasing in the trap. 

Here, we demonstrate 20 seconds of coherence in an atom interferometer held in an optical lattice, overcoming trap dephasing by using an optical cavity as a spatial mode-filter. After 20 seconds, sensitivity to vibrations is suppressed by $10^3-10^4$ relative to traditional atomic gravimeters at the same sensitivity, due to the continuous accumulation of free evolution phase in the trapped wave packets. Trapping the interferometer allows for the sensitivity to be increased by extending interrogation times rather than wavepacket separations or free fall distances, reducing experimental complexity and potentially minimizing systematics. 

\begin{figure}
\centering
\includegraphics[width = 3.35in]{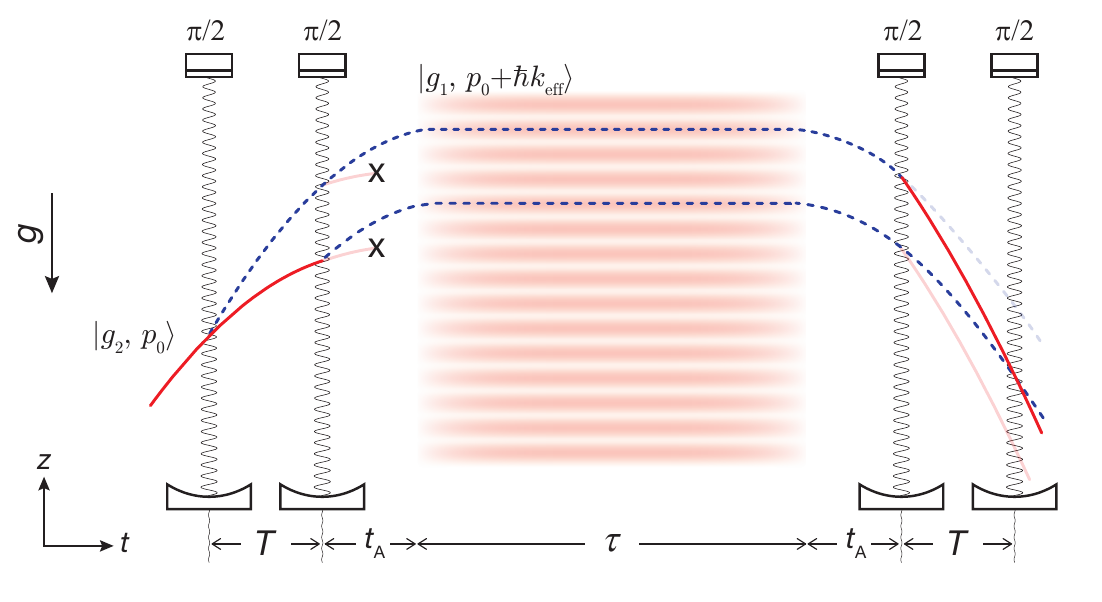} 
\caption{Lattice interferometer schematic. Trajectories of atoms in the lattice interferometer. The red solid lines show trajectories for the state $|g_2, p_0\rangle$, where atoms with momentum $p_0$ are in the state $|g_2\rangle$; the blue dashed lines show the state $|g_1, p_0 + \hbar {\bf k}_{\text{eff}}\rangle$, where atoms kicked by two photons of momentum $\hbar {\bf k}_{\text{eff}}$ are in the state $|g_1\rangle$, where $|g_1\rangle$ and $|g_2\rangle$ are the hyperfine ground states of cesium (typically corresponding to $F=3$ and $F=4$, respectively). Each pulse pair is separated by a time $T$. Between the $\pi$/2 pulses and the lattice hold, atoms move in free fall towards the apex of their trajectory for a time $t_\text{A}$. At the apex, atoms are loaded in a far-detuned optical lattice formed by the mode an optical cavity (red stripes) and remain held in the lattice for a time $\tau$.}
\label{fig1}
\end{figure}

Our matter-wave interferometer builds upon the setup described previously in \cite{Hamilton2015-yp,Jaffe2017-ow}. Cesium atoms are laser-cooled to $\sim$300 nK, prepared in the magnetically-insensitive $m_F$=0 state, and launched millimeters upwards into free fall (see Methods for details). In free fall, counter-propagating laser beams in the cavity manipulate the atomic trajectories. We stimulate two-photon Raman transitions between the hyperfine ground states of cesium, $F=3$ and $F=4$, imparting two photons' momenta to the atom with each laser pulse. The pulse intensities are tuned to kick atoms with 50\% probability (“$\pi$/2-pulses”), enacting coherent matter-wave beamsplitters which separate the two partial wavepackets with a relative velocity of 2$v_\text{rec}$ = 7 mm/s, where $v_\text{rec}$ is the recoil velocity of a cesium atom absorbing a photon on its D2 line.

\begin{figure*}
\centering
\includegraphics[width = 6.5in]{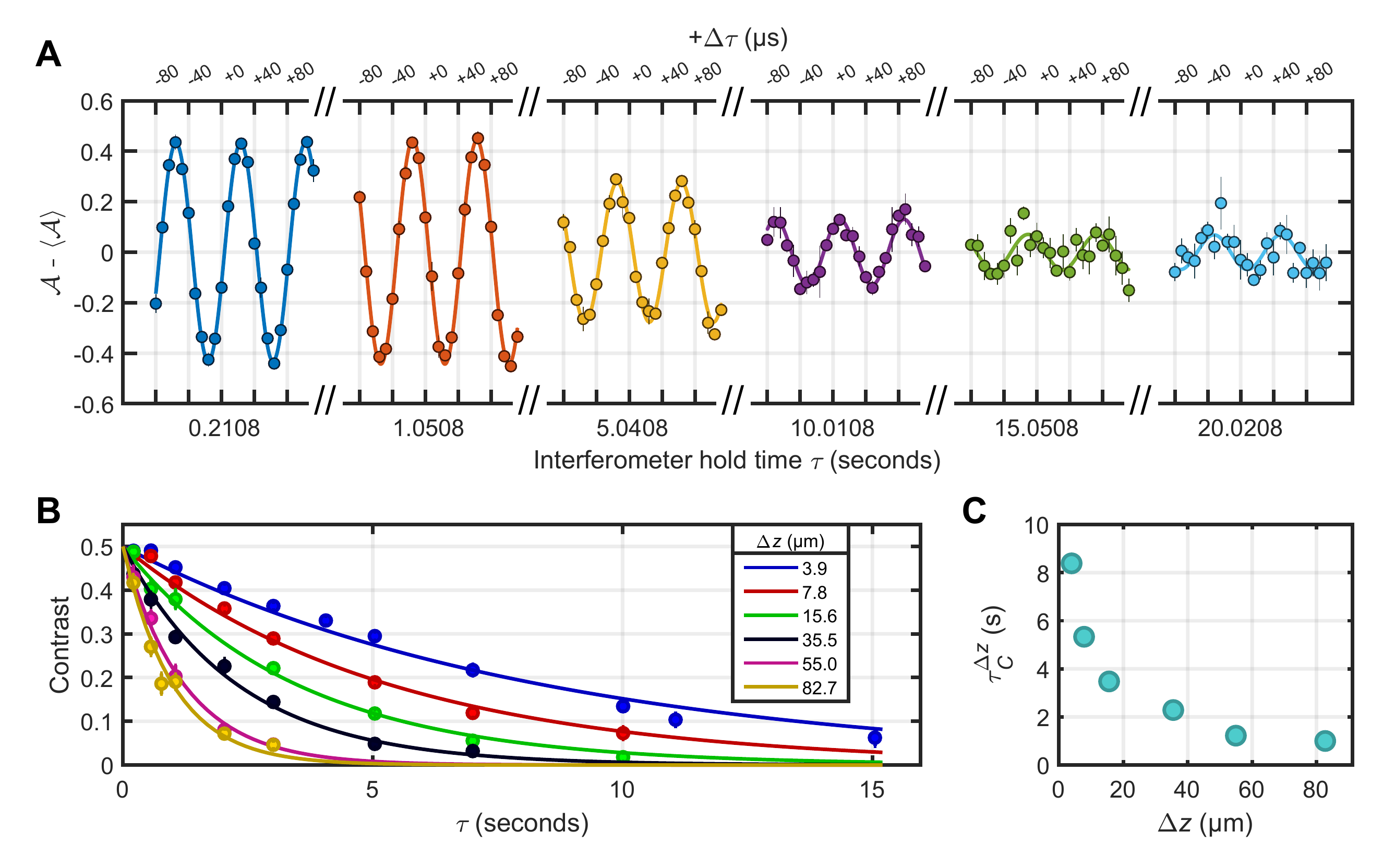} 
\caption{Interference fringes and contrast decay. (A) Interference fringes are visible after holding the atoms for up to $\tau$ = 20 seconds ($\sim$28,300 Bloch oscillations) in an optical lattice. Each data point (filled circles) is averaged over several interferometer cycles. Error bars show the 1-$\sigma$ spread. We fit each fringe to a sine function (solid lines) where the fitted amplitude gives the fringe contrast, $C$. For each hold time, the mean asymmetry $\langle \mathcal{A} \rangle$ is removed for clarity. This interferometer used a pulse separation time of $T$ = 0.516 ms and $t_\text{A}$ = 11 ms. The oscillation frequency is $\omega_{\text{FE}}(\Delta n=9) = 2 \pi \times $(12.7 kHz), consistent with the vertical separation of $\Delta z$ = 3.9 $\mu$m, or $\Delta n$ = 9 lattice sites. (B) Contrast measured (filled circles) as a function of hold time $\tau$ and wavepacket separation $\Delta z$. The contrast lifetime $\tau_\text{C}(\Delta z)$ for each wavepacket separation $\Delta z$ is extracted from fits (solid lines) to an exponential decay, $C(t,\tau_\text{C}(\Delta z)) = 0.5 \  e^{-\frac{\tau}{\tau_\text{C}(\Delta z)}}$. (C) Contrast lifetimes $\tau_\text{C}(\Delta z)$ are observed to decrease with increasing wavepacket separation. }
\label{fig2}
\end{figure*}

Our lattice interferometer employs two pairs of $\pi$/2-pulses, as shown in Figure 1. The first pair, separated by a time $T$, splits the matter-waves into four paths. We select two of these paths, in which the wavepackets are vertically separated by a distance $\Delta z = 2 v_\text{rec} T$ while sharing the same hyperfine ground state and external momentum. At the apex, we adiabatically load atoms into the ground band of a far-detuned optical lattice with a period of $d = \lambda_\text{latt}/2$, where the laser wavelength is $\lambda_\text{latt}$ = 866 nm. The atoms are suspended in the lattice for a time $\tau$, undergoing Bloch oscillations \cite{Bloch1929-aa,Zener_Clarence1934-xt,Ben_Dahan_M1996-ao} with period $\tau_\text{B} = (m_\text{Cs} g d / h )^{-1} \approx 707.5\ \mu$s, where $m_\text{Cs}$ is the atomic mass of cesium, and $g$ is the local gravitational acceleration. Next, we adiabatically unload atoms from the lattice and recombine the wavepackets using the final pair of $\pi$/2-pulses. At the last pulse, the atomic matter-waves interfere according to the phase difference $\Delta \phi = \phi_\text{upper} - \phi_\text{lower}$ accumulated between the upper and lower arms. As a result, the probabilities $P_{3,4}$ of detecting an atom in the output ports corresponding to $F=3$ and $F=4$ oscillate with this phase difference $P_{3,4} = \frac{1}{2}\left[1 \pm C \cos(\Delta\phi)\right]$, where $C$ is the fringe contrast. Since only two of the four spatially unresolved output ports interfere, our maximum contrast is 0.5 in this geometry. We measure the atom number ($N_{3,4}$) in each port through fluorescence imaging and extract the total interferometer phase $\Delta \phi$ from oscillations in the population asymmetry $\mathcal{A} = \frac{N_3-N_4}{N_3+N_4} = C \cos(\Delta \phi)$.

\begin{figure*}
\centering
\includegraphics[width = 6in]{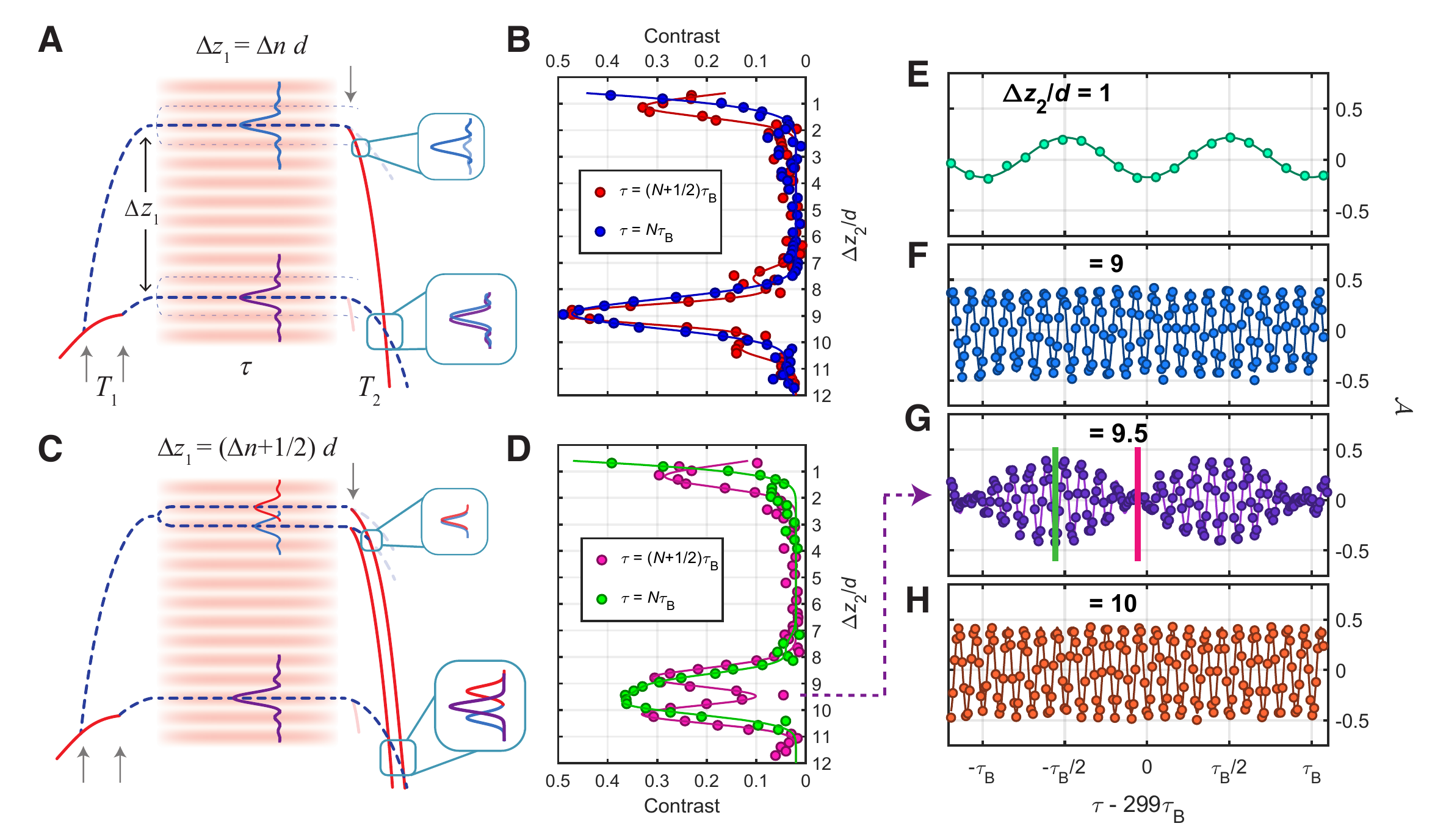} 
\caption{Lattice delocalization. (A) Space-time trajectory for integer lattice loading, with $\Delta z_1/d$ = 9. (B) Contrast as a function of $\Delta z_2$ for the integer lattice loading shown in (A), after holding for integer $N$ (blue) or half-integer $N+1/2$ (red) Bloch periods, $\tau_\text{B}$. For each $\Delta z_2$, the fringe contrast was obtained by varying $\Delta \phi_\text{L}$. (C) Space-time trajectory for half-integer lattice loading, with $\Delta z_1/d$ = 9.5. The upper and lower arms acquire different spatial distributions. (D) Contrast as a function of $\Delta z_2$ for the half-integer lattice loading shown in (C), after holding for integer $N$ (green) or half-integer $N+1/2$ (pink) Bloch periods. E-H) Free evolution fringes from the half-integer lattice loading $\Delta z_1/d$ = 9.5 in (C), near a hold time of $\tau \sim 299 \ \tau_\text{B}$ (0.2115 s). The oscillation frequencies verify the discretized arm separations that result from loading a half-integer initial wavepacket separation. For $\Delta z_2/d$ = 1, 9, and 10, panels (E, F, H) show oscillation frequencies corresponding to the labelled arm separations of $\omega_{\text{FE}}$($\Delta n$=1), $\omega_{\text{FE}}$($\Delta n$=9), and $\omega_{\text{FE}}$($\Delta n$=10), respectively. (G) At $\Delta z_2/d$ = 9.5, oscillations at $\omega_{\text{FE}}$($\Delta n$=9), and $\omega_{\text{FE}}$($\Delta n$=10) add, showing a beating interference in the free evolution phase at their difference frequency $\omega_{\text{FE}}$($\Delta n$=1).  }
\label{fig3}
\end{figure*}

For traditional atomic gravimeters operating in free space, the total phase $\Delta \phi = \Delta \phi_\text{L} + \Delta \phi_\text{FE}$ is dominated by the atom-light interaction phase $\Delta \phi_\text{L}$ ("laser phase"), while the free evolution phase $\Delta \phi_\text{FE}$ is zero. Each laser pulse contributes a phase $\phi_i$ proportional to the atoms' position. For atoms in free fall, $\Delta \phi_\text{L}$ can provide a sensitive measurement of accelerations such as gravity \cite{Peters1999-rn}, which influence the atoms’ position at each laser pulse.

This lattice-based interferometer realizes nearly the opposite scenario in that the free evolution phase $\Delta \phi_\text{FE}$ constitutes more than 99\% of the total phase alongside only a small contribution from $\Delta \phi_\text{L}$. In this pulse sequence, $\Delta \phi_\text{L}$ = $(\phi_1 - \phi_2) - (\phi_3 - \phi_4)$ \cite{Charriere2012-sa,Zhang2016-dl}. Between laser pulses, each arm accumulates a phase $\phi_\text{FE}$ which can be calculated by integrating the Lagrangian $\mathcal{L}$ over the classical trajectory \cite{Storey1994-cl}. Suspending the interferometer causes a phase difference
\begin{equation}
    \Delta \phi_\text{FE} = \frac{1}{\hbar} \int_{\tau} \Delta U dt = m_\text{Cs} g \Delta z \ \tau / \hbar, 
\end{equation}
to accumulate between the upper and lower arms during the lattice hold due to the gravitational potential energy difference, $\Delta U = m_\text{Cs} g \Delta z$, across a vertical separation $\Delta z$. There is zero net contribution to the free evolution phase outside of the lattice hold.

Figure 2A shows interference fringes due to the gravitational potential energy difference from a vertical separation of $\Delta z = 3.9\ \mu$m, corresponding to nine lattice spacings. Fringes remain visible as the interferometer is trapped for up to $\tau$ = 20 seconds, at which point $\Delta \phi_\text{FE}$ = 1.6 Mrad. Without the lattice to hold atoms against Earth’s gravity, interrogating atoms in free fall for 20 seconds would require a vacuum system about 0.5 km tall. In this interferometer, atoms travelled less than 2 millimeters. This allows highly sensitive yet very compact atomic setups, which help suppress spatially-dependent systematic effects such as gravitational and magnetic field gradients. Moreover, measuring $\Delta \phi_\text{FE}$ by increasing $\tau$ allows for the substantial suppression of systematics which are independent of the hold time. 

Interferometer sensitivity increases with longer hold times $\tau$ and larger wavepacket separations $\Delta z$, with a signal-to-noise proportional to the contrast. At the smallest separation ($\Delta z = 3.9\ \mu$m), we measure a 1/$e$ contrast lifetime in the lattice of $\tau_\text{C}(\Delta z) = 8.4(4)$ seconds (Fig. 2B). The contrast lifetimes decrease for larger wavepacket separations (Fig. 2C), presumably from residual imperfections of the cavity-filtered optical lattice beam. 

\begin{figure*}
\centering
\includegraphics[width = 5in]{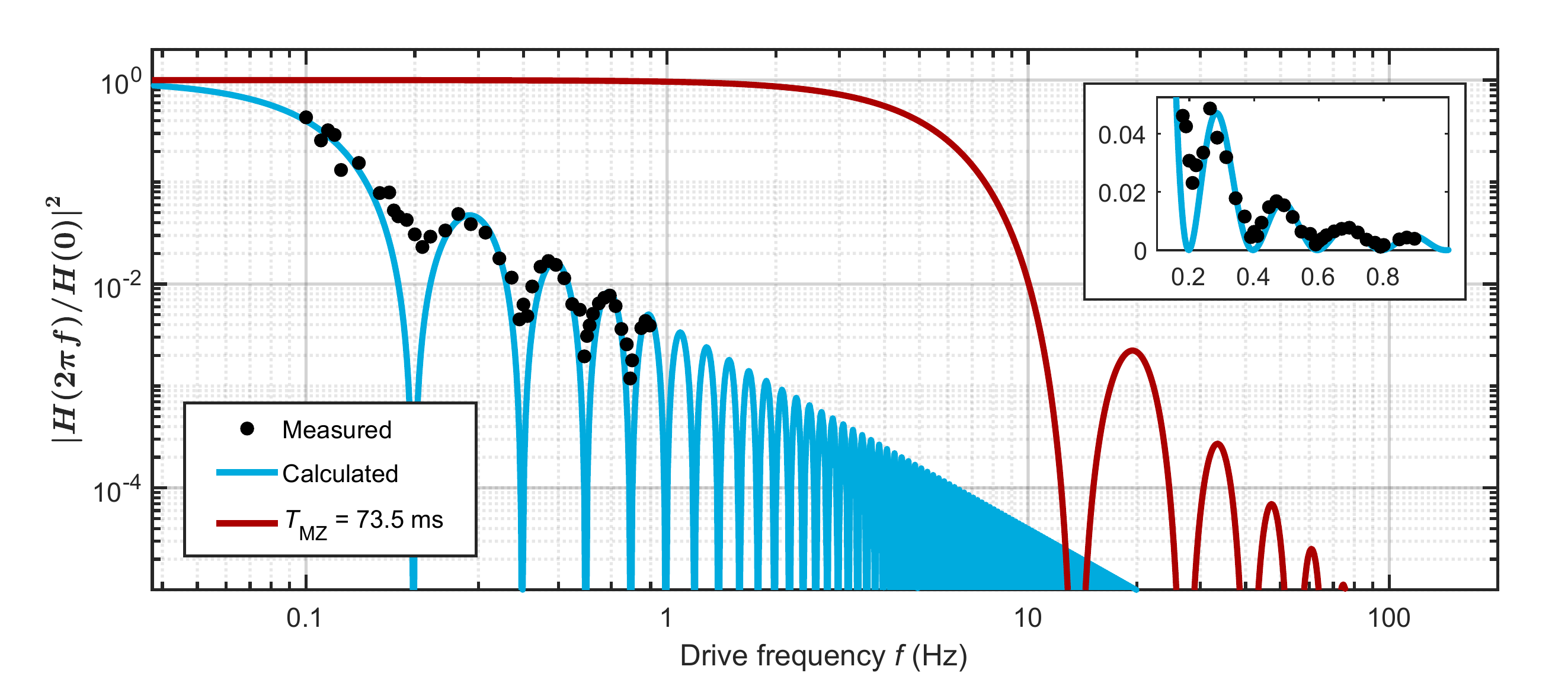} 
\caption{Lattice interferometer acceleration transfer function. Measured (black circles) and calculated (blue line) transfer function, normalized to 1 at dc as
$\left|H\left(2\pi f \right)\right|^2 / \left|H\left(0\right)\right|^2$, for a lattice interferometer with $T$ = 1.066 ms, and $\tau$ = 5.0 seconds. A Mach-Zehnder interferometer with comparable dc acceleration sensitivity has a pulse separation time of $T_{\text{MZ}}$ = 73.5 ms, whose transfer function is plotted in red. The lattice interferometer transfer function suppresses the phase noise arising from mechanical vibrations in the critical $\sim$0.1-100 Hz frequency range by 2-3 orders of magnitude over the equivalent Mach-Zehnder interferometer. Inset: Data and calculations are plotted on a linear scale and show good agreement. Each data point represents the mid-fringe phase variance from $\sim$70 experimental runs, in response to an applied acceleration which pushes the vacuum chamber at a drive frequency $f$. The transfer function measurement is detailed in Methods. }
\label{fig4}
\end{figure*}

In general, the optical lattice modifies the spatial structure of the atomic wavepackets. Operating the lattice interferometer requires an understanding of these dynamics. In a typical atom interferometer, contrast can be observed as long as the wavepacket positions at the time of the final pulse are within the coherence length of the sample, as determined by the thermal de Broglie wavelength $\lambda_\text{T}$ of the atomic wavepacket. In our setup, this condition is met when the distance travelled by the partial wavepackets after the third pulse $\Delta z_2 = 2 v_\text{rec} T_2$ is within $\lambda_\text{T}$ of the initial wavepacket separation $\Delta z_1 = 2 v_\text{rec} T_1$, where $T_1$ and $T_2$ refer to the pulse separation times before and after the lattice hold, respectively (see Fig. 3A). When matching the free-space wavepacket separations $\Delta z_{1,2}$ to an integer number $\Delta n$ of lattice spacings, contrast is observed within $\lambda_\text{T}$ of $\Delta z_2 = \Delta z_1$. This can be seen in Fig. 3B by the peak in contrast around $\Delta z_2 = \Delta z_1 = 9d$.

The finite spatial extent of the atomic wavepacket can coherently distribute across multiple adjacent lattice sites, leading them to acquire a finer spatial substructure at the lattice spacing $d$. For atoms with $\lambda_\text{T} \geq d$, the partial wavepacket along each arm (see Fig. 3A) necessarily distributes into adjacent lattice sites, effectively creating additional interferometer arms with a vertical separation of $\Delta  n=1$. We verify coherence between the added paths in two ways. First, we measure a peak in the contrast envelope near $\Delta z_2 \approx d$, showing that components of the wavefunction separated by one lattice site interfere. Accordingly, Figures 3B and 3D show spatial overlap within $\lambda_\text{T}$ of $\Delta z_2 = d$ for various choices of $\Delta z_1$. Second, we vary $\tau$ to obtain the free evolution fringe shown in Fig. 3E. The oscillation frequency, $\omega_\text{FE}(\Delta n) = \left( \frac{m_\text{Cs} g d}{\hbar} \right) \Delta n = \left(\frac{2\pi}{\tau_\text{B}} \right) \Delta n$ with $\Delta n=1$, verifies that $\Delta \phi_\text{FE}$ accumulates from the gravitational potential energy difference across a vertical arm separation of $\Delta z = d$.

When the initial separation is a non-integer number of lattice spacings $\Delta z_1 \neq \Delta nd$, lattice loading can cause the wavepackets’ spatial distributions to differ between the upper and lower arms. Figure 3C shows an example with $\Delta z_1 = 9.5 d$: one arm splits between two adjacent lattice sites, becoming two new interferometer arms which are separated by $\Delta n = 1$ from one another, and $\Delta n$ = 9 and 10 from the distant arm. We verify coherence by changing $\Delta z_2$ to interfere different combinations of the three arms; Figure 3D shows the corresponding contrast envelopes within $\lambda_\text{T}$ of $\Delta z_2/d$ = 1, 9, and 10. Moreover, we vary $\tau$ to obtain free evolution fringes for each combination of arms (see Fig. 3 E, F and H, respectively). The observed oscillation frequencies $\omega_\text{FE}(\Delta n)$ show the quantization of arm separations in multiples of the lattice period $d$.

At $\Delta z_2/d$ = 9.5, the lower arm is partially spatially overlapped with both upper arms, and the final pulse closes the two interferometers with separations of $\Delta n$ = 9 and 10 simultaneously. As a result, we observe the adjacent wavepackets separated by $\Delta n$ = 1 coming into and out of phase at their difference frequency $\omega_\text{FE}(\Delta n=1) $, while interfering with the distant arm constructively after integer Bloch periods $\tau = N \tau_\text{B}$, or destructively after half-integer Bloch periods $\tau = (N+1/2) \tau_\text{B}$. This results in a beating interference in the free evolution phase, which we observe by varying $\tau$ (see Fig. 3G).

Reaching seconds of phase coherence in the lattice allows this trapped interferometer to overcome the limiting noise source in state-of-the-art atomic gravimeters: phase noise from vibrations \cite{Hu2013-xv,Freier2016-qg}. Instead of reducing the vibrations themselves, seconds of hold time cut off the interferometer phase sensitivity to vibration noise by orders of magnitude across the problematic $\sim$0.1-100 Hz range, where it is difficult to suppress \cite{Hensley1999-dr}. For instance, hold times of $\tau$ = 20 seconds cause the vibration noise sensitivity to drop rapidly as the frequency rises above 10 mHz, suppressing the phase sensitivity to vibrations by 3-4 orders of magnitude across the critical $\sim$0.1-100 Hz band. In Methods, we derive the interferometer phase response to the acceleration noise caused by vibrations, i.e. the acceleration transfer function $H(\omega)$, accounting for vibration sensitivity of both $\Delta \phi_\text{L}$ and $\Delta \phi_\text{FE}$.

We directly measure $\left|H\left(2\pi f\right)\right|^2$ by using a voice coil to apply accelerations to the vacuum chamber at a frequency $f$, and record the mid-fringe phase variance as the drive frequency $f$ is varied. Figure 4 shows the agreement between the measured and calculated transfer functions, $\left|H\left(2\pi f\right)\right|^2$/$\left|H\left(0\right)\right|^2$, for a lattice interferometer with $T$ = 1.066 ms and $\tau$ = 5.0 seconds. This measurement confirms that $\left|H\left(2\pi f\right)\right|^2$ oscillates at a frequency commensurate with the hold time, $f_\text{0}^{\text{latt}}$ = 1/$\tau$ = 0.2 Hz. For comparable dc acceleration sensitivity to this lattice interferometer, a Mach-Zehnder (MZ) interferometer would use a pulse separation time of $T_{\text{MZ}}$ = 73.5 ms, setting the first zero in its transfer function at $f_\text{0}^{\text{MZ}}$ = 13.6 Hz, nearly 70x higher than $f_\text{0}^{\text{latt}}$. For a target dc sensitivity, the vibration immunity in this lattice geometry can be further enhanced by increasing $\tau$ and decreasing $T$.

Overall, the lattice interferometer realizes an attractive scheme for metrology by holding atoms to directly probe the potential energy difference, rather than dropping atoms to measure accelerations. This approach strongly suppresses vibration noise while extending interrogation times in a compact volume, overcoming the major limitations in conventional atomic gravimetry. This lattice geometry is therefore well-suited for precision gravimetry \cite{Freier2016-qg}, with exciting prospects for geophysics \cite{Wu2019-oi}, and fundamental tests of short-ranged forces such as dark energy \cite{Jaffe2017-ow}, Casimir forces \cite{Harber2005-jq}, or short-ranged gravity \cite{Kapner2007-il}. Additionally, measuring the phase due to a potential without subjecting the atoms to an acceleration represents a milestone towards observing a gravitational analogue of the Aharonov-Bohm effect \cite{Hohensee2012-ed}, which can provide a novel measurement of Newton's constant $G$ through the gravitational potential.

\begin{acknowledgments}
We thank Philipp Haslinger for collaboration in the lab, Zachary Pagel and Paul Hamilton for valuable discussions, and Weicheng Zhong for experimental assistance. This material is based upon work supported by the David and Lucile Packard Foundation, the National Science Foundation under grant No. 1708160 as well as the National Aeronautics and Space Administration Grants 1629914 and 1612859. 
\end{acknowledgments}

\bibliographystyle{apsrev4-1}
\bibliography{latt_ai_arxiv}

\beginsupplement
\section*{Methods}

\subsection*{Experimental sequence}%
\subsubsection*{Atom source}%
An experimental cycle begins by using a two-dimensional magneto-optical trap (2D-MOT) to load cesium atoms into a 3D-MOT near the center of our in-vacuum optical cavity, cooling the ensemble to sub-Doppler temperatures $\sim 10$ $\mu$K. The atoms are then loaded into a 3D optical lattice and Raman sideband cooling is performed to lower the ensemble temperature to near the recoil limit ($< 300$ nK). This final stage of laser cooling leaves atoms in the $|F=3,m_F=3 \rangle$ stretched state. Upon release from the lattice, microwave pulses transfer the ensemble into the magnetically-insensitive $|F=3,m_F=0\rangle$ state. 

Atoms are then loaded into the ground band of an optical lattice, where the lattice potential is formed by 852 nm light resonant to a TEM$_{00}$ mode of the optical cavity. We perform a lattice launch, sending about $\sim$10\% of the atoms upwards into free fall. For the experiments shown in the main text, launched atoms reached the apex of their trajectory after 20 ms of free fall time (corresponding to 1.96 mm). After the launch, a Doppler-sensitive Raman $\pi$-pulse (Gaussian temporal profile, 130 $\mu$s FWHM) is applied to narrow the Doppler spread of the atomic ensemble to $\sim$8 kHz before interferometry.

\subsubsection*{Interferometry}

The first pair of $\pi/2$ laser pulses are applied as atoms move upwards towards the apex, where the first pulse separation time $T_1$ determines the spatial separation of the atomic wavepacket before the lattice hold. Before the hold, a resonant "blow-away" laser pulse selects the two paths corresponding to atoms in the state $|g_2,p_0+\hbar k_{\text{eff}}\rangle$ to load into the lattice. Atoms are typically held in the $|g_2\rangle = |F=3, m_F = 0\rangle$ hyperfine ground state. After unloading the atoms from the lattice, another blow-away laser pulse removes atoms that have scattered lattice photons and decayed to $|F=4\rangle$. Atoms that have scattered but decay into $|F=3\rangle$ can remain trapped and contribute to contrast loss. 

The last pair of interferometry pulses occur as atoms move downwards after release from the lattice hold, and their pulse separation time $T_2$ determines the separation of the paths being interfered at the last pulse. We extract $\Delta\phi$ by computing the population asymmetry, $\mathcal{A}$, which we use to normalize for fluctuations in the total atom number. We fit the fringes to a sine function, whose amplitude $C$ gives the fitted fringe contrast. 

For adiabatic lattice loading (unloading), the lattice intensity ramps up (down) over 350 $\mu$s. The lattice uses a wavelength of $\lambda_{\text{latt}}$ = 866 nm, which is the wavelength red-detuned of the D2 line that minimizes the scattering-per-recoil-depth when considering both the D1 and D2 lines, that still fits within the cavity mirror reflectivities. The lattice power is intensity-stabilized to minimize atom heating effects due to fluctuating laser power.

\subsubsection*{Detection}

We detect the number of atoms in each interferometer output port using fluorescence imaging. Immediately after the end of the interferometry sequence, we apply a laser pulse which resonantly pushes atoms in the $\left|F=4\right\rangle $ ground state away from atoms in the $\left|F=3\right\rangle $ state, spatially separating the two interferometer output ports. Once the clouds are sufficiently far apart, we use MOT and repump light (resonant with the $\left|F=4\right\rangle \rightarrow \left|F'=5\right\rangle$ and $\left|F=3\right\rangle \rightarrow \left|F'=4\right\rangle$ transitions, respectively) to image both output ports, collecting fluorescence from the atoms for about 15~ms on a CCD camera.

The near-resonant light which excites the atoms also induces a large background signal on the camera caused by scattering off of other components in our system. In order to isolate the fluorescence signal from the atoms, we first collect the image $I_{fore}$ including the atoms and the background, and then after a delay of 40 ms in which the atoms have fallen away we take a second image $I_{back}$ which contains only background light. In the bulk of this work, we directly extract the atomic signal $I_{atom}=I_{fore}-I_{back}$ by subtracting the background image from the foreground. However, when using this approach, we find that the uncertainty in $I_{atom}$ is dominated by the fluctuation of the background light during the 40 ms gap between collecting the two images. This uncertainty in the atomic fluorescence image dominates the uncertainty in the  population asymmetry $A$, which is calculated from $I_{atom}$.

Interference fringes are readily visible with standard subtraction of the background image for hold times up to 15 seconds. When the participating atom number drops below $\sim$50,000 atoms however, the imaging noise from background light can become almost comparable to the contrast.

In Figure 2A of the main text, we calculate the atomic signal $\tilde{I}_{atom}=I_{fore}-\tilde{I}_{back}$
using a more accurate estimate $\tilde{I}_{back}$ of the true background signal constructed by using the information in $I_{fore}$ which is outside of the regions containing atoms; our approach is similar to that discussed in chapter four of Ref.~\cite{kronjagerthesis} in the context of removing fringes from absorption images. Our procedure is detailed in the next paragraph; in brief, we begin by calculating an orthonormal set of basis images $\tilde{B}_n$ which represent the typical spatial structures observed in the background images. Then, for each measured image $I_{fore}$, we construct $\tilde{I}_{back}$ by adding together the basis images with weights determined by their overlap with the structures in $I_{fore}$ outside of the regions containing the atom clouds. This scheme relies on the fact that the spatial correlation in the background light between different regions of the image is stronger than the temporal correlation in the background light over the 40 ms delay. 

Our procedure for constructing the background estimates $\tilde{I}_{back}$ proceeds as follows. Note that each image may be thought of as a vector whose length is the number of pixels in the image, such that standard techniques of linear algebra are applicable. First, we perform a principal component analysis (PCA) on a large set of average-subtracted background images, which produces an orthonormal set of basis images $B_{n}$, sorted in ascending order by the amount of variance in the set of background images which occurs along each basis image. In addition to the basis elements produced by PCA, we prepend an additional element representing the pixel-wise average across all the background images as the element $B_{0}$.
We next apply a Gram-Schmidt orthonormalization procedure to produce a modified set of basis images $\tilde{B_{n}}$ which are orthogonal in the region which does not contain the atom clouds, satisfying the condition $\left(\tilde{B_{n}}\circ M\right)\cdot\left(\tilde{B_{m}}\circ M\right)=\delta_{mn}$,
where the mask $M$ is defined as,

\[
M=\begin{cases}
1 & \mathrm{\mathrm{region\,without\,atoms}}\\
0 & \mathrm{regions\,containing\,atom\,clouds}
\end{cases}
\]
Here, the symbol $\cdot$ denotes the inner product and the symbol $\circ$ denotes element-wise multplication (the Hadamard product). After constructing this basis, for each image $I_{fore}^{j}$ we calculate the overlaps $O_{n}^{j}=\text{\ensuremath{\left(\tilde{B_{n}}\circ M\right)\cdot\left(I_{fore}^{j}\circ M\right)}}$
between the image and the first 35 basis elements in the region which does not contain atoms. We next construct the background estimate $\tilde{I}_{back}^{j}=\sum_{n=0}^{34}O_{n}^{j}\tilde{B}_{n}$, which includes the estimated background in the regions containing atoms.
Finally, for each image $j$ we calculate the atomic signal $\tilde{I}_{atom}^{j}=I_{fore}^{j}-\tilde{I}_{back}^{j}$. We find that the use of $\tilde{I}_{atom}$ typically reduces the excess noise caused by background light fluctuations by a factor of $1.6$ relative to $I_{atom}$.

\subsection*{Optical cavity parameters}
The optical cavity consists of a flat, gold-coated mirror (radius of curvature $\text{ROC}_{1} = \infty$) and a curved dielectric mirror ($\text{ROC}_{2} = 10$ m). They are separated by a length $L = 37.4886(2)$ cm, inferred from the measured free spectral range $\text{FSR} = 399.845(6)$ MHz. The length and radii of curvature determine the waist size $w_{0} = 724$ $\mu$m at the lattice laser wavelength $\lambda = 866$ nm. This corresponds to a Rayleigh range $z_{R} = 1.90$ m. The transverse mode spacing is $\delta f_{mn} = 25.1$ MHz. and contributes to the cavity's spatial mode-filtering.

This cavity beam geometry is favorable for lattice interferometer. For the lattice hold, the cavity length is stabilized to the TEM$_{00}$ Gaussian mode, suppressing the spatial influence of higher-order transverse modes by $\sim$8 cavity linewidths $\gamma_{\text{fwhm}}$ (per mode order) to smooth the trap potentials at each lattice site. Furthermore, the near-planar cavity geometry has a Rayleigh range $z_{R}$ much greater than the interferometer arm separation $\Delta z \sim 4-350 \mu m$, allowing the trap potentials experienced by the two arms to be highly homogeneous.

The intensity reflectivities of the mirrors are $R_{1} = 0.98$ and $R_{2} = 0.973$ for the gold and dielectric mirrors, respectively, at $\lambda_{\text{latt}} = 866$ nm. The second is implied relative to the first (which is approximately known) via the measured linewidth (full width at half maximum) $\gamma_{\text{fwhm}} = 3.03(2)$ MHz. The finesse is therefore $\mathcal{F} = 132$. This resonant power enhancement results in $\sim$13 mW of input 866 nm light providing a lattice depth of $\sim$6 E$_{\text{rec}}$ for atoms at the center of the cavity mode.
More details about the cavity can be found in \cite{Hamilton2015-yp,jaffethesis}.

\subsection*{Lattice laser frequency stabilization}
Our basic laser lock scheme is described in detail in \cite{Hamilton2015-yp}. The length of a ``transfer" cavity on the optical table serves as a frequency reference for all lasers which enter the in-vacuum ``science" cavity. This includes the lattice laser (866 nm), the interferometry laser (852 nm), and a ``tracer" laser (780 nm). ($\leq$ 2 $\mu$W) of tracer light is used to stabilize the length of the science cavity, while the other lasers which interact more strongly with the cesium atoms are off. The length of the transfer cavity is stabilized in turn to a ``reference" laser, which is locked to a cesium transition through modulation transfer spectroscopy.

The lattice laser is frequency stabilized to the transfer cavity using a Pound-Drever-Hall (PDH) scheme. A free-space electro-optic modulator (EOM) phase modulates the light to create sidebands at $f_{\text{PDH}} \sim 25$ MHz. This light is sent through a fiber EOM, driven with a tunable high frequency ($f_{\text{offset}}\sim$ 0.6-1.2 GHz). This makes copies of the PDH error signal, separated by $f_{\text{offset}}$. 
These first-order manifolds are tunable via $f_{\text{offset}}$. Locking these manifolds to the transfer cavity allows the lattice laser to be tuned into resonance with the science cavity.

\subsection*{Cavity vertical alignment to gravity}
The radial trap frequency in the 1D lattice is low ($\approx$ 3 Hz), so even a small projection of Earth's gravity along the radial direction would cause atoms to leak out the sides of the trap. To avoid this, the tilt of the optical cavity is stabilized to within $<$ 100 $\mu$rad of Earth’s gravitational pull. The tilt stabilization feedback has been described previously (see the supplemental information in \cite{Jaffe2017-ow} for details), and uses an electronic tiltmeter signal to feed back to air pads supporting the optical table. The setpoint along each axis is determined by maximizing the atom number in a 6 second lattice hold.

\subsection*{Lattice interferometer vibration sensitivity analysis}

Phase noise from vibrations has been studied for the traditional Mach-Zehnder geometry, where all of the measured phase $\Delta\phi$ comes from the laser phase $\Delta\phi_{\text{L}}$. We expand this formalism to quantify the effect of vibration noise in our interferometer geometry, where the free evolution phase $\Delta\phi_{\text{FE}}$ has a significant contribution.

In traditional interferometers, the position of the mirror that is used to retro-reflect the interferometer light sets the inertial reference frame of the measurement. In the cavity interferometer however, the position of both cavity mirrors sets the inertial frame. In considering vibration noise in our apparatus, the cavity can be treated as a rigid body since the cavity length is stabilized with a feedback bandwidth of $\sim$45 kHz, much faster than the typical frequencies of mechanical vibrations. 
To analyze the interferometer phase noise caused by cavity vibrations, we derive the transfer function from acceleration noise of the cavity to phase noise in the lattice interferometer, for both the laser phase $\Delta\phi_{\text{L}}$ and free evolution phase $\Delta\phi_{\text{FE}}$.

\subsubsection*{Lattice interferometer laser phase}
The acceleration transfer function relating vibrations of the cavity to noise in $\Delta\phi_{\text{L}}$ is derived here. To focus this analysis on the phase noise from mechanical vibrations, which impact primarily the low frequency band of $\sim$1-100 Hz, the laser pulses are assumed to occur instantaneously (i.e. with zero pulse duration).
As discussed in the main text, the total lattice interferometer laser phase $\Delta\phi_{\text{L}}$ resulting from the four $\pi/2$ pulses is 
\begin{equation}\label{Dphi_laser_total}
    \Delta\phi_{\text{L}} = (\phi_1 - \phi_2) - (\phi_3 - \phi_4).
\end{equation}

Consider a phase jump $d\phi_{\text{L}}$ which occurs at time $t_{\text{jump}}$. A jump in the laser phase results from e.g. vibrations changing the position of the cavity with respect to the atoms. This jump in $d\phi_{\text{L}}$ shifts the total interferometer phase by $d\phi$. These phase shifts $d\phi$ and $d\phi_{\text{L}}$ are related by the sensitivity function $g_{\text{L}}(t)$, which is defined as
\begin{equation}\label{dphi_laser}
    d\phi = g_{\text{L}}(t) d\phi_{\text{L}}
\end{equation}
From Eq. \ref{Dphi_laser_total}, if the phase jump occurs during between pulses 1 and 2, the total interferometer phase shifts by $d\phi=-d\phi_{\text{L}}$. Similarly, if the phase jump occurs between the pulses 3 and 4, the interferometer phase shifts by the opposite amount $d\phi=+d\phi_{\text{L}}$. Phase jumps $d\phi_{\text{L}}$ which occur during $T'$ (between pulses 2 and 3) do not produce an overall phase shift in the interferometer. 
This gives the following sensitivity function for the laser phase,
\begin{figure}[H]
 \begin{minipage}{.4\textwidth}
    \begin{equation}\label{g_L}
    g_{\text{L}}(t) = \begin{cases} 
              -1, & 0<t<T \\
              0, & T < t < T+T' \\
              1, & T+T' < t < 2T+T' \\
              0, & \text{else}
           \end{cases}
    \end{equation}
  \end{minipage}
  \begin{minipage}{.55\textwidth}
    \centering
    \includegraphics[width=7cm, trim = {0 1cm 0 0.5cm}]{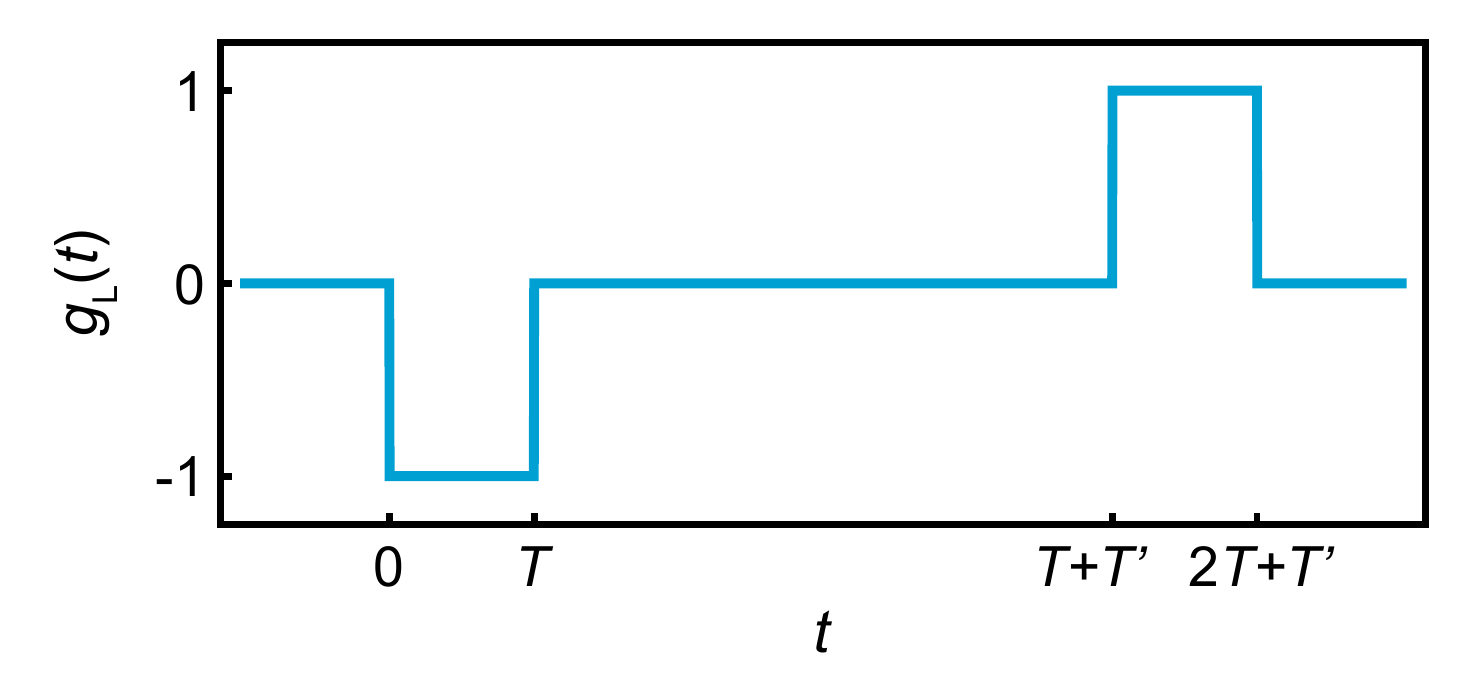}
    \caption{Laser phase sensitivity function, $g_{\text{L}}(t)$}
  \end{minipage}%
\end{figure}
\noindent where the first interferometer pulse occurs at $t=0$.

Integrating the laser phase noise during the interferometer $d \phi_{\text{L}}(t)$ against $g_{\text{L}}(t)$ gives the total interferometer phase fluctuation $\delta\phi$,
\begin{equation}
    \delta\phi = \int g_{\text{L}}(t) d\phi_{\text{L}}(t) = \int^{\infty}_{-\infty} g_{\text{L}}(t) \frac{d\phi_{\text{L}}(t)}{dt} dt
\end{equation}
Instead of analyzing noise in the time-domain, it is better to characterize the frequency components of the noise in the Fourier domain, where this expression can be re-written in terms of the Fourier transforms (indicated with a tilde) as,
\begin{equation}\label{dPhiLtoL}
    \delta\phi = \int d\omega \ H_{\text{L}}^{\phi_{\text{L}}}(\omega) \ \tilde{\phi}_{\text{L}}(-\omega)
\end{equation}
where $H_{\text{L}}^{\phi_{\text{L}}}(\omega)=-i\omega \tilde{g}_{\text{L}}(\omega)$ is defined as the transfer function from laser phase noise $\tilde{\phi}_{\text{L}}(\omega)$ to interferometer phase noise $\delta\phi$. The Fourier transform of 
$g_{\text{L}}(t)$ (Eq. \ref{g_L}) is given by 
\begin{equation}
    \tilde{g}_{\text{L}}(\omega) = \int^{\infty}_{-\infty} e^{-i\omega t} g_{\text{L}}(t) dt = \frac{4}{i\omega}e^{-\frac{i}{2}\omega(2T+T')} \sin{\left(\frac{\omega T}{2}\right)} \sin{\left(\frac{\omega (T+T')}{2}\right)}
\end{equation}

To compare with free evolution phase noise, the laser phase transfer function can be re-written in terms of the acceleration noise caused by vibrations of the cavity. The laser phase noise due to vibrations $\phi_{\text{L}}(t)$ can be expressed in terms of position noise,
\begin{equation}
    \phi_{\text{L}}(t) = k_{\text{eff}}(x_{\text{cavity}}(t)-x_{\text{atom}}(t))
\end{equation}
To convert to acceleration noise, consider a Fourier decomposition of the position noise where a perturbation at frequency $\omega$ changes the relative position of the cavity mirrors and the atoms by $\tilde{x}(\omega)$. Taking the second time-derivative gives an expression for the position fluctuations in terms of accelerations, $\tilde{x}(\omega) = -\frac{1}{\omega^2} \tilde{a}(\omega)$. The laser phase noise now becomes,
\begin{equation}
    \tilde{\phi}_{\text{L}}(\omega) = -\frac{k_{\text{eff}}}{\omega^2} \tilde{a}(\omega)
\end{equation}
in which $\tilde{a}(\omega)$ is the acceleration noise of the cavity due to vibrations. This form of $\tilde{\phi}_{\text{L}}(\omega)$ can now be used in Eq. \ref{dPhiLtoL} to define a new transfer function, which converts from acceleration noise to interferometer phase noise by
\begin{equation}
    H_{\text{L}}^{a}(\omega) = -\frac{k_{\text{eff}}}{\omega^2} H_{\text{L}}^{\phi_{\text{L}}}(\omega)
\end{equation}

The variance of the interferometer laser phase $(\sigma_{\text{L}})^2$ can be calculated (see Refs. \cite{Cheinet2008, jaffethesis}) by integrating the acceleration transfer function for the laser phase $|H_{\text{L}}^{a}(\omega)|^2$ against the acceleration noise power spectral density $S_{a}(\omega)$, 
\begin{equation}
    (\sigma_{\text{L}})^2 = \int^{\infty}_{0} d\omega |H_{\text{L}}^{a}(\omega)|^2 S_{a}(\omega)
\end{equation}
where the norm-squared of the acceleration transfer function for $\Delta\phi_{\text{L}}$ is
\begin{equation}
|H_{\text{L}}^{a}(\omega)|^2 = \frac{16 k_{\text{eff}}^{2}}{\omega^{4}}\sin^{2}\left( \frac{\omega T}{2}\right) \sin^{2}\left(\frac{\omega(T+T')}{2} \right).
\end{equation}

\subsubsection*{Lattice interferometer free evolution phase}
We now derive the transfer function from the acceleration noise of the cavity to free evolution phase noise. As discussed in the main text (see Eq. 2), this interferometer accumulates $\Delta\phi_{\text{FE}}$ from the linear gravitational potential gradient across the interferometer, $\Delta U=mg \Delta z$, such that 
\begin{equation}
    \Delta\phi_{\text{FE}} = \frac{1}{\hbar} \int_\tau dt \Delta U = \frac{1}{\hbar} \int_\tau dt \left( m g \Delta z  \right)
\end{equation}
Here, $\Delta z$ is the vertical separation during the lattice hold. This separation is enforced by the interferometer geometry to be 
\begin{equation} \label{eq:dz}
\Delta z = \Delta n d \approx 2 v_{\text{rec}} T = \frac{\hbar k_{\text{eff}}}{m} T
\end{equation}
where $\Delta n$ is an integer number of lattice sites. 

To extend the analysis of vibrations to $\Delta\phi_{\text{FE}}$,
we can assume that the atomic wavepackets follow the vibrating lattice. Cavity vibrations can be treated as the acceleration noise $a(t)$ experienced by atoms trapped in the vibrating lattice. Acceleration is a useful physical quantity here because it allows us to invoke the equivalence principle to equate this noise $a(t)$ to a noisy gravitational field $g \to g+a(t)$, for which the noisy potential energy entering the interferometer can be expressed as
\begin{equation}
    \Delta U = m \ a(t) \ \Delta z
\end{equation}
%

A sensitivity function $g_{\text{FE}}(t)$ for the free evolution phase can then be defined via the relation
\begin{equation} \label{eq:dphi_FE}
    d\phi(t) = g_{\text{FE}}(t) \left[ m \ a(t) \ \Delta z(t) / \hbar  \right] dt,
\end{equation}
where $d\phi(t)$ is the amount of free evolution phase accumulated due to the noisy acceleration $a(t)$ in the interval $dt$. The additional factors set $g_{\text{FE}}$ to be unitless.

Vibrations of the cavity are coupled to the atoms while they are trapped in the lattice. Therefore, an acceleration caused by a vibration of the cavity will only shift $\Delta\phi_{\text{FE}}$ if it occurs during $\tau$. The free evolution phase sensitivity function is therefore
\begin{figure}[H] %
 \begin{minipage}{.34\textwidth}
    \begin{equation} \label{gFE}
        g_{\text{FE}}(t) = 
            \begin{cases} 
                1, & t \in \tau \\
                0, & \text{else}
            \end{cases}
    \end{equation}
  \end{minipage}%
  \begin{minipage}{.6\textwidth}
    \centering
     \includegraphics[width=7cm, trim = {0 1cm 0 .5cm}] {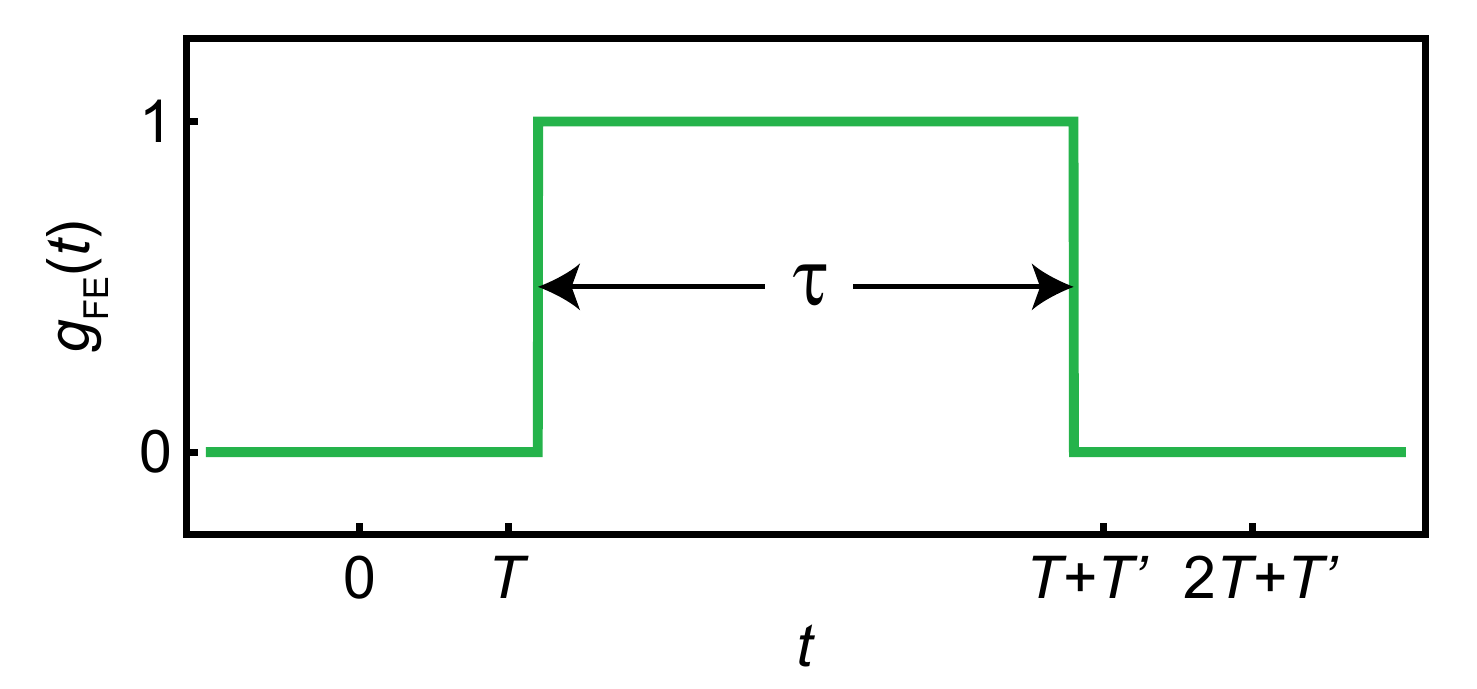}
    \caption{Free evolution phase sensitivity function, $g_{\text{FE}}(t)$ }
  \end{minipage}
\end{figure} 
\noindent 

As with the laser phase, the vibration noise contribution $\delta \phi$ to the free evolution phase is calculated by integrating eq. \ref{eq:dphi_FE} over the interferometer duration. Since $g_{\text{FE}}$ is nonzero only during the lattice hold, we can substitute in (from Equation \ref{eq:dz}) $\Delta z = \hbar k_{\text{eff}} T / m$, giving

\begin{align}
    \delta\phi &= k_{\text{eff}} T \int dt \ g_{\text{FE}}(t) \ a(t) =  k_{\text{eff}} T \int d\omega \ \tilde{g}_{\text{FE}}(\omega) \ \tilde{a} (-\omega) \nonumber \\
    & = \int d\omega \ H^{a}_{\text{FE}}(\omega) \tilde{a}(-\omega).
\end{align}

\noindent
In the above equation, tildes again represent a Fourier transform and $H_{\text{FE}}^{a}(\omega) =  k_{\text{eff}} T \tilde{g}_{\text{FE}}(\omega)$ has been defined as the transfer function from acceleration noise to free evolution phase noise. The Fourier transform of $g_{\text{FE}}(t)$ is

\begin{equation}
    \tilde{g}_{\text{FE}}(\omega) = \frac{2}{\omega} e^{-i\omega(T+\tau)} \sin{\left(\frac{\omega \tau}{2}\right)}
\end{equation}

The free evolution phase variance $(\sigma_{\text{FE}})^2$ is again calculated by integrating against the acceleration noise power spectral density
\begin{equation} \label{eq:sigma_equals_H_U}
    (\sigma_{\text{FE}})^2 = \int^{\infty}_{0} d\omega |H_{\text{FE}}^{a}(\omega)|^2 S_{a}(\omega)
\end{equation}
\noindent 
where the norm-squared of the acceleration transfer function for $\Delta\phi_{\text{FE}}$ is 
\begin{equation}
    |H_{\text{FE}}^{a}(\omega)|^2 = \frac{ 4 k_{\text{eff}}^{2}T^{2} }{ \omega^{2}} \sin^{2}\left( \frac{\omega \tau}{2}\right).
\end{equation}

\subsubsection*{Comparison}
Table \ref{table:vib_sensitivity} compares the transfer functions for the lattice interferometer laser phase and free evolution phase $\left( |H_{\text{L}}^{a}(\omega)|^2,|H_{\text{FE}}^{a}(\omega)|^2 \right)$,
along with that of the Mach-Zehnder (MZ) laser phase $|H_{\text{MZ}}^{a}(\omega)|^2$. The MZ laser phase is given by $\Delta\phi_{\text{MZ}}^{\text{L}} = \phi_{1} - 2\phi_{2} + \phi_{3}$, where $\phi_{i}$ is the laser phase for each of the three pulses in the geometry. Each of the pulses are separated by a duration $T$. The sensitivity function is thus \cite{Cheinet2008}
\begin{figure}[H] %
 \begin{minipage}{.34\textwidth}
    \begin{equation}
    g_{\text{MZ}}(t) = \begin{cases} 
              -1, & 0 < t < T \\
              1,  & T < t < 2T \\
              0,  & \text{else}
           \end{cases}
    \end{equation}
  \end{minipage}%
  \begin{minipage}{.6\textwidth}
    \centering
    \includegraphics[width=7cm, trim = {0 1cm 0 .5cm}] {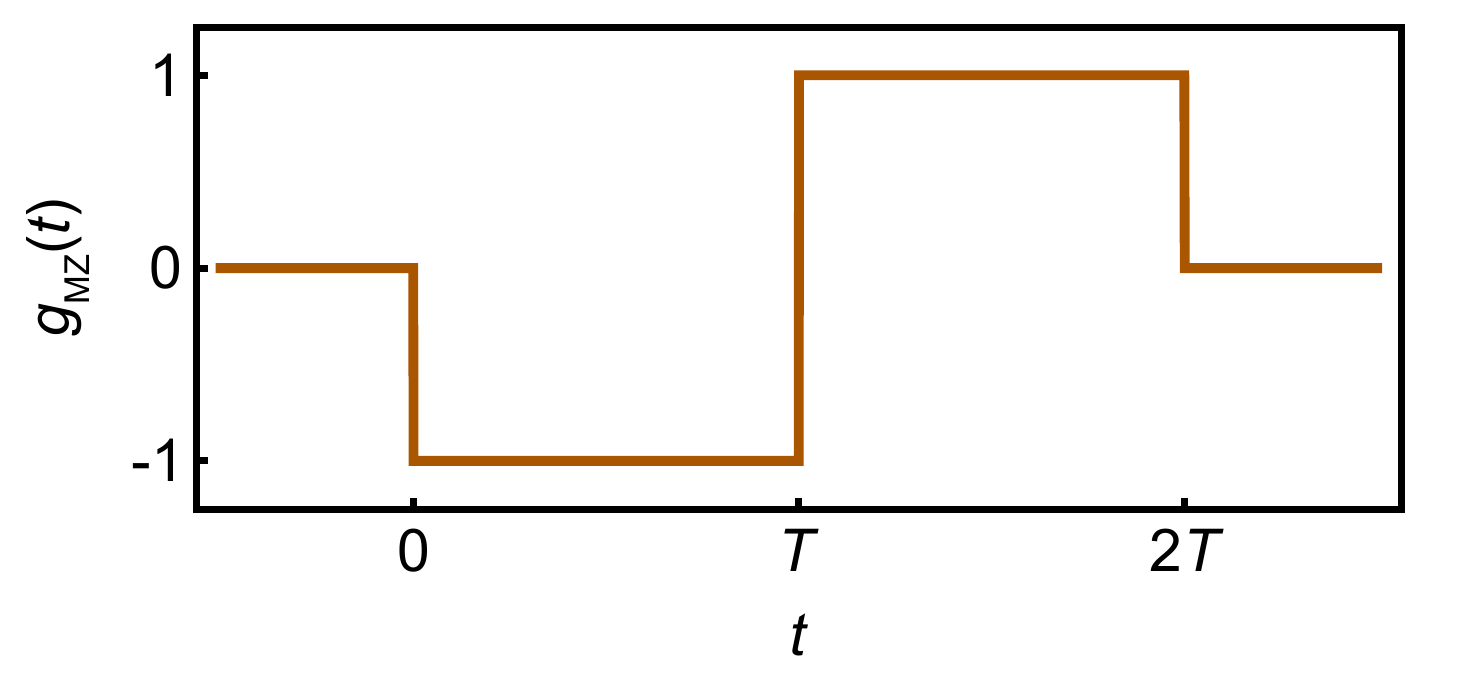}
    \caption{Mach-Zehnder laser phase sensitivity function, $g_\text{MZ}(t)$}
  \end{minipage}
\end{figure} 
%

\renewcommand\theadfont{\bfseries}
\renewcommand{\arraystretch}{2.25}
\newcolumntype{Z}{ >{\centering\arraybackslash} X }
\newcolumntype{Q}{ >{\centering\arraybackslash} p{0.20\textwidth} }

\noindent
\begin{table}[H]
\caption[Vibration-sensitivity expressions]{Vibration-sensitivity expressions for relevant interferometer phases. These phases are discussed in terms of a noisy variable $X$, and a function $f$ that maps $X$ to a common variable, acceleration $a$. Other notation in the table follows that of the derivations in the text.}
\label{table:vib_sensitivity}
\begin{tabularx}{\textwidth}{
		|p{0.15\textwidth}
		|Q
		|Z
		|Z|
	}
	\hline
	 	& \thead{Mach-Zehnder\\laser phase $\phi_{\text{MZ}}^{\text{L}}$} 
	 	& \thead{Lattice AI\\laser phase $\phi_{\text{latt}}^{\text{L}}$}
	 	& \thead{Free evolution\\phase $\phi_{\text{latt}}^{\text{FE}}$} \\
	\hline
	\vspace{-8mm} \bm{$\phi_{\text{AI}}$} 		& \vspace{-8mm} $\phi_{1} - 2\phi_{2} + \phi_{3}$ 
												& \vspace{-8mm} $(\phi_{1} - \phi_{2}) - (\phi_{3} - \phi_{4})$ 
												& \vspace{-8mm} $\frac{m}{\hbar} \int \!\! dt \,a(t) \Delta z(t)$ \\ \hline
	\vspace{-8mm} \textbf{Noisy }$\bm{X(t)}$	& \vspace{-8mm} $\phi_{\text{L}}(t)$ & \vspace{-8mm} $\phi_{\text{L}}(t)$ & \vspace{-8mm} $a(t)$ \\ \hline
	\vspace{-8mm} \bm{$\Delta \phi_{i}(X)$}		& \vspace{-8mm} $\int \!\! dt \,g(t) \frac{dX}{dt}$ & \vspace{-8mm} $\int \!\! dt \,g(t) \frac{dX}{dt}$ 
												& \vspace{-8mm} $k_{\text{eff}} T\int \!\! dt \, g(t) X(t)$ \\ \hline
	\multirow{2}{*}[0.1cm]{\vspace{-0mm}\bm{$g(t)$}} 
												&\multirow{2}{*}[0.1cm]{
													\begin{minipage}{0.95\linewidth}
														\hspace{-2mm}
														\includegraphics[width=\linewidth]{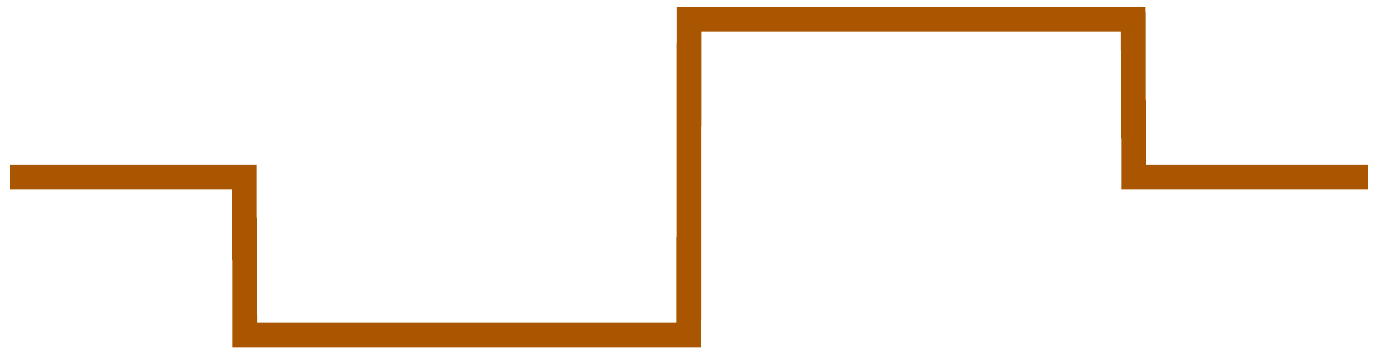}
													\end{minipage}
													}
												&\multirow{2}{*}[0.1cm]{
													\begin{minipage}{\linewidth}
														\includegraphics[width=\linewidth]{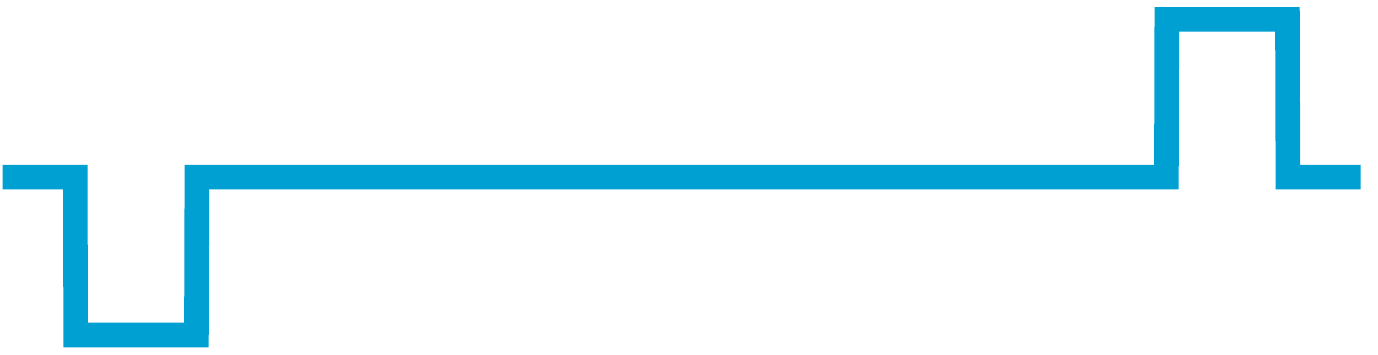}
													\end{minipage}
												}
												&\multirow{2}{*}[0.1cm]{
													\begin{minipage}{\linewidth}
														\includegraphics[width=\linewidth]{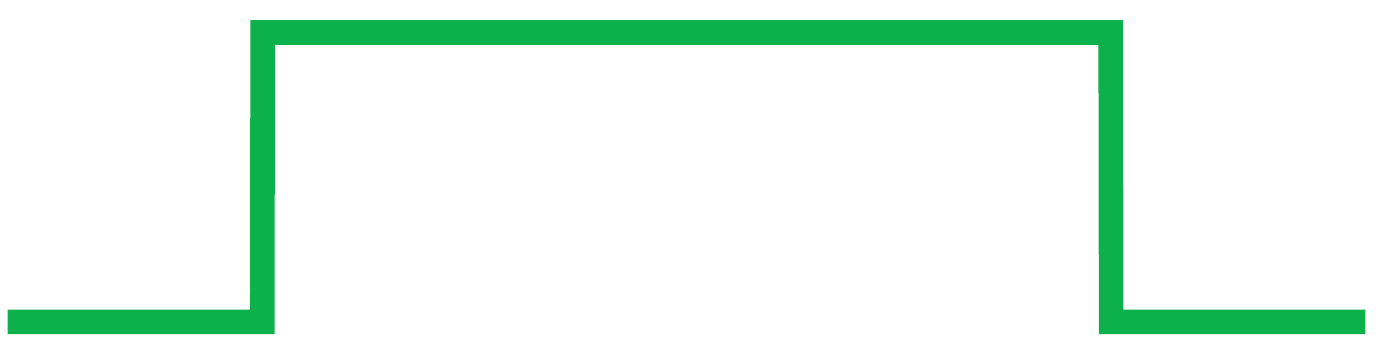}
													\end{minipage}
												}\\
	&&&\\ \hline 
%
	\vspace{-8mm} \bm{$f: X \mapsto a$} 		& \vspace{-8mm} $-\frac{k_{\text{eff}}}{\omega^{2}}$
												&\vspace{-8mm}  $-\frac{k_{\text{eff}}}{\omega^{2}}$
												& \vspace{-8mm}  1 \\ \hline
	\multirow{2}{*}[0.1cm]{\vspace{-0mm}\bm{$|H^{a}_{i}(\omega)|^{2}$}} 
			&\multirow{2}{*}[0.1cm]{	$\frac{16 k_{\text{eff}}^{2}}{\omega^{4}} \sin^{4}\left( \frac{\omega T}{2} \right)$					}
			&\multicolumn{1}{l|}{\multirow{2}{*}[0.1cm]{ \fontsize{10}{4}
							$\frac{16 k_{\text{eff}}^{2}}{\omega^{4}}\sin^{2}\left( \frac{\omega T}{2}\right) \sin^{2}\left(\frac{\omega(T+T')}{2} \right)$}}
			&\multirow{2}{*}[0.1cm]{	$\frac{4 k_{\text{eff}}^{2}T^{2}}{\omega^{2}}\sin^{2}\left( \frac{\omega \tau}{2}\right)$	}
			\\
	&&&\\ \hline 
\end{tabularx}
\end{table}

\subsection*{Transfer function measurement details}
Figure 4 of the main text demonstrates a measurement of the lattice interferometer transfer function. A force $F_{\text{vc}}$ at frequency $f$ was applied to the vacuum chamber using a voice coil. This results in interferometer phase noise discussed above, with an acceleration noise power spectral density $S_{a}(\omega) \propto a_{0} \delta(\omega - 2 \pi f_{0})$. Measuring the variance of the phase noise in this way gives the transfer function $|H_{\text{latt}}^{a}(2\pi f_{0})|^{2}$.

The interferometer phase noise for a given voice coil drive frequency $f$ is read out by running $\sim$ 70 shots of the experiment sitting at mid-fringe, where the output port population ratio is linear in phase. The resulting population ratio spread is then proportional to the phase noise spread. However, the envelope of $|H_{\text{latt}}^{a}(\omega)|^{2}$ varies by about two orders of magnitude over the measurement's frequency range. On the low-sensitivity end, measuring the interferometer phase noise due to vibrations is limited by imaging noise. On the high end, the measured population ratio is no longer linear with interferometer phase, and the phase spread is difficult to infer from the population ratio spread. As a result, we use a fed-forward drive amplitude to keep the resulting phase spread between these two limits, as described below.

The vacuum chamber is supported against gravity on seismic attenuation air pads with a resonance frequency $f_{\text{res}} = 2.7$ Hz. The applied force $F_{\text{vc}}$ pushes against the spring force from the air pads, defining a new equilibrium position $z_{\text{eq}}$ for the sum of gravity plus these two spring forces. All data was taken with $f<1$ Hz, sufficiently below $f_{\text{res}}$ that the transfer function of the air pads is approximately flat. That is, we can consider the motion of the vacuum chamber to track the equilibrium position $z_{\text{eq}}$.

A voice coil force at frequency $f$ is linearly proportional to an applied current $I(t) = I_{0}\sin(2 \pi f t)$. This implies a motion of the vacuum chamber $z(t) = z_{0} + \delta z \sin(2 \pi f t)$, where $\delta z$ is the amplitude of the position displacement, and is proportional to $I_{0}$. The resulting acceleration is
\[
    a(t) = -\frac{a_{0}'}{\left( 2 \pi f \right)^{2}} \sin(2 \pi f t) := a_{0}(f) \sin(2 \pi f t)
\]

While $\delta z$ does not depend on $f$, $a_{0}(f)$ does. We thus feed-forward a factor proportional to $\left( 2 \pi f \right)^{2}$ to the amplitude of the drive current. That is, the current that drives the voice coil is

\[
    I(t) = \left(  2 \pi f \right)^{2} I_{0} \sin( 2 \pi f t)
\]

This gives an experimentally measurable phase variance across the full frequency range. We then correct for the extra factor of $\left( 2 \pi f \right)^{2}$ in the drive to extract the transfer function $|H_{\text{latt}}^{a}(\omega)|^{2}|$ and to plot it in Fig. 4. The overall amplitude of the data is fitted to the normalized transfer function; this is the only free parameter.

\end{document}